\documentclass{emulateapj}
\usepackage{natbib}
\usepackage{epsfig} 
\usepackage{graphicx} 
\usepackage{subfigure}
\usepackage{float}
\usepackage{amsmath}
\usepackage{color} 
\usepackage{amssymb} 
\usepackage{amsfonts}
\usepackage{units}
\usepackage{mathtools}
\usepackage{bm}
\usepackage[colorlinks,linkcolor=blue,anchorcolor=green,citecolor=blue]{hyperref}
\usepackage{ulem}
\def\be{\begin{equation}}
\def\ee{\end{equation}}

\shorttitle{Magnetar coherent radio emission}
\shortauthors{Wang et al.}
\begin{document}

\title{Coherent radio emission from a twisted magnetosphere after magnetar-quake}
\author{Weiyang Wang\altaffilmark{1,2,3}, Bing Zhang\altaffilmark{4}, Xuelei Chen\altaffilmark{1,2,5}, Renxin Xu\altaffilmark{3,6}}\email{wywang@bao.ac.cn}
\affil{$^1$Key Laboratory for Computational Astrophysics, National Astronomical Observatories, Chinese Academy of Sciences, 20A Datun Road, Beijing 100101, China}
\affil{$^2$University of Chinese Academy of Sciences, Beijing 100049, China}
\affil{$^3$School of Physics and State Key Laboratory of Nuclear Physics and Technology, Peking University, Beijing 100871, China}
\affil{$^4$Department of Physics and Astronomy, University of Nevada, Las Vegas, NV 89154, USA}\email{zhang@physics.unlv.edu}
\affil{$^5$Center for High Energy Physics, Peking University, Beijing 100871, China}
\affil{$^6$Department of Astronomy, School of Physics, Peking University, Beijing 100871, China}

\begin{abstract}
Magnetars are a class of highly magnetized, slowly rotating neutron stars, only a small fraction of which exhibit radio emission.
We propose that the coherent radio curvature emission is generated by net charge fluctuations from a twist-current-carrying bundle (the j-bundle) in the scenario of magnetar-quake. 
Two-photon pair production is triggered, which requires a threshold voltage not too much higher than $10^9$ V in the current-carrying bundle, and which can be regarded as the ``open field lines'' of a magnetar. Continued untwisting of the magnetosphere maintains change fluctuations, and hence coherent radio emission, in the progressively shrinking j-bundle, which lasts for years until the radio beam is too small to be detected. 
The modeled peak flux of radio emission and the flat spectrum are generally consistent with the observations.
We show that this time-dependent, conal-beam, radiative model can interpret the variable radio pulsation behaviors and the evolution of the X-ray hot spot of the radio transient magnetar XTE J1810$-$197 and the high-$B$ pulsar/AXP PSR J1622$-$4950.
Radio emission with luminosity of $\lesssim10^{31}\,{\rm erg\,s^{-1}}$ and high-frequency oscillations are expected to be detected for a magnetar after an X-ray outburst.
Differences of radio emission between magnetars and ordinary pulsars are discussed.
\end{abstract}


\section{Introduction}
Magnetars are highly magnetized and slowly rotating neutron stars (NSs), which are historically identified as two related classes, anomalous X-ray pulsars (AXPs) and soft gamma-ray repeaters (SGRs; see \citealt{kas17} for a review).
They exhibit dramatically variable X-ray and $\gamma$-ray emissions including short bursts, large outbursts, giant flares and quasi-periodic oscillations (QPO), which are believed to be powered by the dissipation of their enormous internal magnetic fields, typically $10^{14}-10^{16}$ G \citep{dun92}.
Even their ``persistent'' emission is far from being steady. 
Magnetars are often accompanied by glitches which show irregular spin-down evolutions (e.g., \citealt{kas00,dib08,sas14}). 
These behaviors may be related to the origin of magnetar bursts.

The leading scenario of magnetar bursts invokes quakes in the neutron star crust. Within such a scenario, when the pressure induced by the internal magnetic field exceeds a threshold stress, the magnetic energy releases from the crust into the magnetosphere \citep{tho01}, leading to particle acceleration and a short burst of radiation \citep{tho95}. 
It is then expected that the magnetar bursts may exhibit characteristics of self-organized criticality, as has been observed in earthquakes (\citealt{soc}; e.g., \citealt{che96,dun98,gog99}).

There have been many attempts to detect radio pulsations from magnetars (e.g., \citealt{gae01,bur06}).
However, among the known 23 magnetars, only four have been identified as pulsed radio emitters, with the addition of one more high-B pulsar, PSR J1119$-$6127, which might be a magnetar (\citealt{ola14} for a review, \citealt{arc16,gog16}).
The radio emission mechanism of magnetars seems to be different from that of ordinary pulsars.
For instance, AXP XTE J1810$-$197 was a switched-on radio-transient during 2003 to 2009.
The radio emission appeared following an X-ray outburst, and then decayed with the X-ray emission abating \citep{cam16}.
1E 1547.0$-$5408, another AXP, was also identified as a switched-on radio-transient intermittently following its 2009 outburst \citep{bur09}.
Their radio emission differs from that of ordinary rotation-powered pulsars by having extremely variable flux densities, flatter spectra, and pulse profiles (e.g.,\citealt{cam07b,laz08}).
Similar distinct characteristics are also found in other radio magnetars \citep{cam07c,lev10,sha13}.
The correlation between the radio and X-ray emission indicates that the radio emission is likely powered by quake-triggered currents in the magnetosphere rather than by the steady spindown power.

The coherent radio emission mechanism for radio pulsars is poorly understood due to their high brightness temperatures.
That for magnetars is even more so thanks to their peculiar environment and the different observational properties from the ordinary pulsars.
It was proposed that the coherent radio counterparts of the X-ray bursts from the magnetars have fluxes as high as 1 kJy, which is reminiscent of the solar type III radio bursts \citep{lyu02,lyu06}.
However, such a predicted radio flux is much higher than the observed peak fluxes for some radio transients (e.g., \citealt{cam16}).
Alternatively, \cite{lin15} argued that stellar oscillations can provide additional voltage in the polar cap region, making a ``magnetar" re-active by crossing the pulsar radio emission death line.
If the radio emission is originated from the oscillation-induced unipolar induction from the open field line regions, the number of observed radio magnetars would be small because of the narrow open field beam size of slow rotators.

In this paper, a coherent curvature radiative model of magnetar-quake-induced net charge fluctuation in the twist-current-carrying bundle is proposed to explain the radio emission of magnetars.
The paper is organized as follows.
An introductory description of magnetar-quake-induced, twisted, and oscillating magnetosphere is presented in Section 2. A radiative model is introduced in Section 3. In Section 4, we simulate the radio pulse profile evolution as magnetosphere untwisting and apply the model to XTE 1810$-$197 and PSR J1622$-$4950, respectively. The results are summarized in Section 5 with some discussion. 
Some detailed calculations are presented in the Appendix.

\section{Magnetar quake-induced twisted magnetosphere}
\subsection{Twisted magnetosphere}

\begin{figure}
\begin{center}
\includegraphics[width=0.48\textwidth]{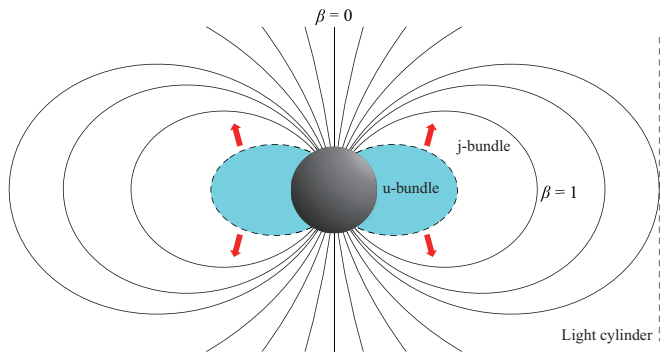}
\caption{\small{ The distribution of $\beta$ in the magnetosphere. The twist angle is adopted as $\Delta\psi=1$. The light blue region is the expanding u-bundle and the dashed lines are the boundaries between the u-bundle and the j-bundle. The solid lines are the electric current that support the twisted magnetic fields.}
\label{fig1}}
\end{center}
\end{figure}

\begin{figure}
\begin{center}
\includegraphics[width=0.48\textwidth]{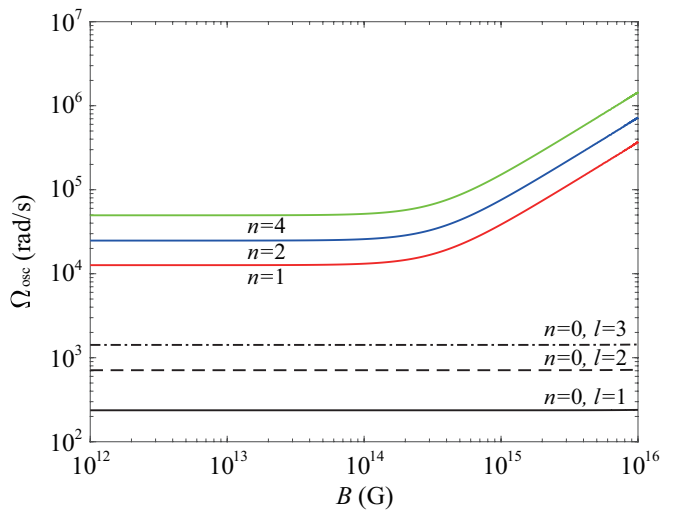}
\caption{\small{Oscillation frequency of the $n=0$\,(black\,lines), $1$\,(red\,line), $2$\,(blue\,line), and $4$\,(green\,line) modes as a function of $B$. The cases of $l=1$\,(black solid\,line), $2$\,(black dashed\,line) and $3$\,(black dotted-dashed\,line) for $n=0$ are presented. Here, the $n>0$ modes are independent of $l$.}}
\label{fig2}
\end{center}
\end{figure}

The twist of magnetic field lines is strong inside the star and vanishes in the magnetosphere \citep{tho02}.
However, during a sudden crust quake, the magnetic field in the outer magnetosphere begins to twist up owing to the magnetic energy release in the crust \citep{bel09}.
Basically, the magnetic field lines are anchored to the star crust and the field geometry is determined by the motion of the footpoints.
The relaxation of the twist makes a network of fractures in the magnetar crust, which leads to the motion of the footpoints, and the entire magnetosphere twists up.
During this process, the ejected current flows along the field lines to the exterior of the star and comes back at other footpoints \citep{tho00}.

The twisted magnetic field is supported by these emerging currents.
This configuration can be assumed to be force-free since the energy release is dominated by the magnetic fields in the magnetosphere \citep{tho02}.
When the starquake shears outer magnetic fields, a toroidal component $B_{\phi}$ develops and the magnetosphere twists up.
In spherical coordinates ($r,\,\theta,\, \phi$), the twist angle can be written as \citep{bel09}
\begin{equation}
\Delta\psi=\int d\phi =\int \frac{B_{\phi}dl}{Br\sin \theta}.
\label{eq2}
\end{equation}
Once the whole magnetosphere is twisted, it cannot be untwisted rapidly because a self-induced electric field builds up, which initiates pair production and accelerates the pairs forming current flows.
Such twist-maintained electric current is given by \citep{bel07},
\begin{equation}
\boldsymbol{j}=\frac{c}{4\pi}\nabla\times\boldsymbol{B}\simeq\frac{c\boldsymbol{B}}{4\pi r}\sin^2\theta\Delta\psi.
\label{eq3}
\end{equation}
The poloidal components $B_{r}$ and $B_{\theta}$ are not much different from the corresponding components for a normal dipole even when twists are strong, e.g., $\Delta\psi\sim1$.
The main difference is in the toroidal term $B_{\phi}$.

To maintain these currents, unsteady pair production in a short timescale is needed.
The minimal charge density that is needed to support the twist-maintained current is
\begin{equation}
\rho_{\rm tw}=\frac{|\boldsymbol{j}|}{c}=\beta \rho_{\rm GJ},
\label{eq4}
\end{equation}
where $\beta$ is a function of $r$ and $\theta$, and $\rho_{\rm GJ}$ is the Goldreich-Julian (GJ) charge density \citep{gol69}, i.e.,
\be
\rho_{\rm GJ}=\frac{-\boldsymbol{\Omega}\cdot \boldsymbol{B}}{2\pi c}\frac{1}{[1-(r^2\Omega^2/c^2)\sin^2\theta]}.
\ee
Here, the distribution of $\beta$ is plotted in Figure \ref{fig1}.
The charge density of the twist-maintained current is a constant for each magnetic field line.
In the corotating frame, the neutral condition of the magnetosphere is $(n_+-n_-)e=\rho_{\rm GJ}$, where $n_+$ and $n_-$ are the density of positrons and electrons respectively, and the current is given by $\boldsymbol{j}=en_+\boldsymbol{v}_+-en_-\boldsymbol{v}_-$.
In the region without pair production, the net charge is zero, and the electric current is maintained purely by the pairs flowing in the closed filed lines, i.e., $(n_+-n_-)\ll n_{\pm}$ \citep{bel13}.
Note that the unstable net charge generation is a necessary condition for coherent emission.
Define
\begin{equation}
u\equiv R\sin^2\theta/r. 
\end{equation} 
In the region where $u \sim 1$, 
it is unlikely to create coherent emission.

The energy of the twisted magnetic field is continually dissipated in the magnetosphere.
It is worth noting that the magnetosphere can be divided into two parts: the twist-maintained current-carrying bundle with twisted field lines (the j-bundle),  and the cavity with untwisted field lines (the u-bundle).
The dissipation is mostly ohmic, since the current is maintained by the electric field $E_{\parallel}$ parallel to $\boldsymbol{B}$.
The dissipation is significant at the transition boundary from the j-bundle to the u-bundle, where the strong currents dissipate as the magnetic field lines twist down rapidly.
As a result, the u-bundle expands from the region of $u=1$  to that of $u=0$ \citep{bel09}, as shown in Figure \ref{fig1}.
The so-called cavity front, $u_b$, is defined as the boundary between the u-bundle and the j-bundle, expanding with an extremely high speed near $u_b = 1$, then decelerates to $u_b\ll1$.
The expansion can be described as \citep{bel09}
\begin{equation}
\frac{du_b}{dt}=-\frac{V}{2BR^2\Delta\psi_0/c+2V't},
\label{eq6}
\end{equation}
where $\Delta\psi_0$ is the initial twist angle, $V$ is the threshold voltage that can trigger plentiful pairs supplying the electric current, and $V'=dV/du$ is the voltage gradient.

 \subsection{Oscillation of the magnetosphere}

In a starquake, the sudden release of energy can create seismic waves.
After a previous large quake (main-quake), smaller quakes (aftershocks) occur as the crust around the displaced fault plane adjusts itself to the effects of the main-quake.
These quakes (main-quake or aftershock) likely excites significant magnetospheric oscillations.
The toroidal modes, preserving the stellar shape, are fundamental modes which are pure shear deformations during stellar oscillations.
Other modes also give rise to bulk compression and vertical motion, which have to do work against the much stronger degeneracy pressure or gravity. 
Therefore, the toroidal modes are most likely excited by starquakes, since the restoring force is due to the Coulomb forces of the crustal ions \citep{dun98}.

The crust is the outermost $\sim$1\,km layer of an NS where ions are locked into a solid lattice. 
It can store much elastic and magnetic energy.
The density scale-height can be neglected for it is typically only a few percent of the crust thickness \citep{cha08}. 
Within the crust, the ions in the solid crust are arranged in a Coulomb lattice whose shear modulus is \citep{str91}
\begin{equation}
\mu= \frac{0.1194}{1+0.595(173/\Gamma)^2}\frac{n_i(Ze)^2}{a},
\label{eq7}
\end{equation}
where $\Gamma=(Ze)^2/(akT)=173$ \citep{far93}, $n_i$ is the number density of ions, $Z$ is the atomic number and $a$ is the lattice constant. 
We adopt a bcc crystal lattice for the crust, with the lattice constant $a=(2/n_i)^{1/3}$.
The number density of ions is $n_i=\rho_i/(Z\mu_em_u)$, where $\rho_i$ is the mass density of the ions, $m_u$ is the atomic mass unit, and $\mu_e$ is the mean molecular weight per electron, which we adopt an intermediate value 2.5 here. 
Then, the shear modulus can be written as
\begin{equation}
\mu=0.09(Ze)^2n_i^{\frac{4}{3}}=9.6\times10^{27}\left(\frac{Z}{32}\right)^{\frac{2}{3}}\left(\frac{\rho_i}{\rho_N}\right)^{\frac{4}{3}}\,\rm{erg\,cm^{-3}},
\label{eq9}
\end{equation}
where $\rho_N=4\times10^{11}\,\rm{cm^{-3}}$ is the neutron drip density.

We consider a spherical coordinate system with $r$ as the radial coordinate.
The toroidal displacement is defined as
\begin{equation}
\boldsymbol{\xi}=\xi_x\boldsymbol{\hat{x}}+\xi_y\boldsymbol{\hat{y}},\qquad\nabla\cdot\boldsymbol{\xi}=0,
\label{eq10}
\end{equation}
where the $x$ and $y$ axes are orthogonal in the plane with ${\hat r}$ as a normal vector.
Within the curst, we assume that the magnetic field $\boldsymbol{B}=B\boldsymbol{\hat r}$ is perpendicular to the curst and constant.
A shear stress tensor for the Lagrangian toroidal displacement is given by \citep{lan70}
\begin{equation}
T_{ij}=\mu\left(\frac{\partial\xi_i}{\partial x_j}+\frac{\partial\xi_j}{\partial x_i}\right),
\end{equation}
where $i=x,y$. For an ion in the crustal lattice, one can obtain
\begin{equation}
\rho_i\frac{\partial^2 \xi_i}{\partial t^2}=\frac{\partial T_{ij}}{\partial x_j}+\frac{1}{4\pi}\left[\left(\nabla\times\delta \boldsymbol{B}+\frac{\partial^2\boldsymbol{\xi}}{c^2\partial t^2}\times \boldsymbol{B}\right)\times \boldsymbol{B}\right]_i,
\label{osc}
\end{equation}
where $\delta \boldsymbol{B}=\nabla\times(\boldsymbol{\xi}\times\boldsymbol{B})$ is the perturbed magnetic field \citep{pir05}.
It is assumed that the solution of equation (\ref{osc}) is $AY_{lm}\rm{exp}\it{[i(k_rr-\omega t)]}$, where $A$ is the amplitude of the displacement and $Y_{lm}$ stand for spherical harmonics. 
In the WKB limit, the vertical wave number are given by $\int k_r dr=n\pi$ \citep{pir05}.
Then, equation (\ref{osc}) can be written as
\begin{equation}
-\rho_i\omega^2 \boldsymbol{\xi}=-\mu\frac{l(l+1)}{R^2} \boldsymbol{\xi}+\frac{\partial}{\partial r}(\mu\frac{\partial\boldsymbol{\xi}}{\partial r})+\frac{B^2}{4\pi}(\frac{\partial^2 \boldsymbol{\xi}}{\partial r^2}+\frac{\omega^2}{c^2}\boldsymbol{\xi}).
\label{eq11}
\end{equation}
One can obtain the solution of equation (\ref{eq11}),
\begin{equation}
\Omega_{\rm{osc}}^2=\frac{v_s^2(k_r^2+k_{\perp}^2)+v_{\rm A}^2k_r^2}{1-v_{\rm A}^2/c^2},
\end{equation}
where $k_{\perp}^2=l(l+1)/R^2$, $v_s=(\mu/\rho_i)^{1/2}$ is the speed of sound, and $v_{\rm A}=B/(4\pi\rho)^{1/2}$ is the Alv{\'e}n speed.
Figure \ref{fig2} shows a few eigenfrequencies with different modes.
These eigenfrequencies are very high so that it is difficult to detect periodic signals in the oscillations.

Because of the anchored field lines, the starquake-induced crust oscillations make the entire magnetosphere oscillating, which drives the fluctuations of the corotating space charges.
The force-free condition for a corotation magnetosphere is 
\begin{equation}
\boldsymbol{E}+\frac{\boldsymbol{v}(r)}{c}\times\boldsymbol{B}=0,
\end{equation}
where $\boldsymbol{v}(r)=\boldsymbol{v}_{\rm spin}+\boldsymbol{v}_{\rm osc}$ is the velocity at $r$ in the magnetosphere, $\boldsymbol{v}_{\rm spin}$ is the spin velocity, and $\boldsymbol{v}_{\rm osc}$ is the oscillation velocity which can be written as \citep{unn89}
\be
\boldsymbol{v}_{\rm osc}=\left(0,\,\frac{1}{\sin\theta}\partial_{\phi}Y_{lm},\,-\partial_{\theta}Y_{lm}\right)\Omega_{\rm osc}A(r){\rm e}^{-i\Omega_{\rm osc} t},
\ee
where $A(r)$ is the oscillation amplitude.
The net space-charge density is given by
\begin{equation}
\rho=\frac{\nabla\cdot\boldsymbol{E}}{4\pi}=\bar\rho+\delta\rho,
\label{eq12}
\end{equation}
where 
\begin{equation}
\begin{split}
\bar\rho=\rho_{\rm GJ}+\frac{\boldsymbol{\Omega}\cdot(\boldsymbol{r}\times \boldsymbol{j})}{c^2[1-(r^2\Omega^2/c^2)\sin^2\theta]},\\
\delta\rho=-\frac{Bl(l+1)Y_{lm}\Omega_{\rm osc}A(r){\rm e}^{-i\Omega_{\rm osc}t}}{4\pi cr[1-(r^2\Omega^2/c^2)\sin^2\theta]}.
\end{split}
\label{eq13}
\end{equation}
The surface displacement amplitude for $m=0$ mode is given by \citep{dun98}
\begin{equation}
A(R)\simeq4.4\times10^2\left(\frac{E_q}{10^{41}\,\rm{erg}}\right)\left(\frac{R}{10\,\rm{km}}\right)^{-1}\left(\frac{M}{1.4\,M_{\sun}}\right)^{1/2}\,\rm{cm},
\label{eqA}
\end{equation}
where $E_q$ is the energy released in the starquake.
For the $m\not=0$ modes, the distribution of strain is difficult, while the amplitude is roughly of the same order of magnitude.
The amplitude at $r$ is estimated to be $A(r)\simeq A(R)r/R$, since the magnetosphere is corotating with the stellar surface.

\section{Radio emission from a magnetar}
\subsection{Pair production}

\begin{figure}
\begin{center}
\includegraphics[width=0.48\textwidth]{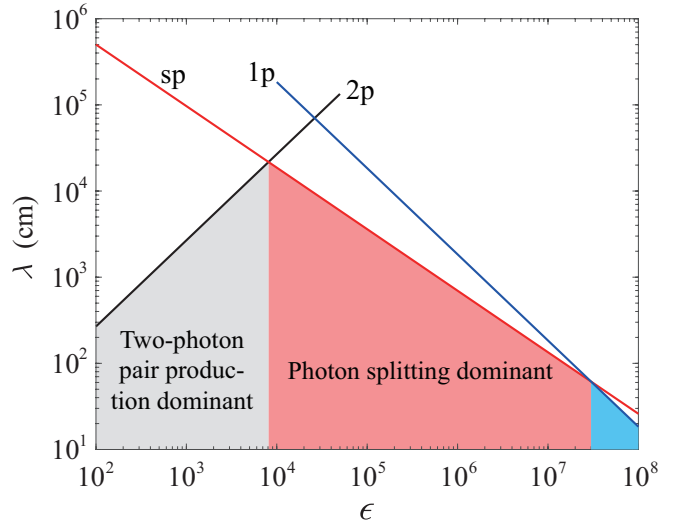}
\caption{\small{The dominant regions of $1\gamma$-, $2\gamma$-pair production and photon splitting at $B=10^{15}\,{\rm G}$ are shown: the black line ($2\gamma$-pair production), red line (photon splitting), and blue line ($1\gamma$-pair production) define the gray region ($2\gamma$-pair production dominant), light red region (photon splitting dominant), and light blue region ($1\gamma$-pair production dominant).}}
\label{fig3}
\end{center}
\end{figure}

\begin{figure}
\begin{center}
\includegraphics[width=0.48\textwidth]{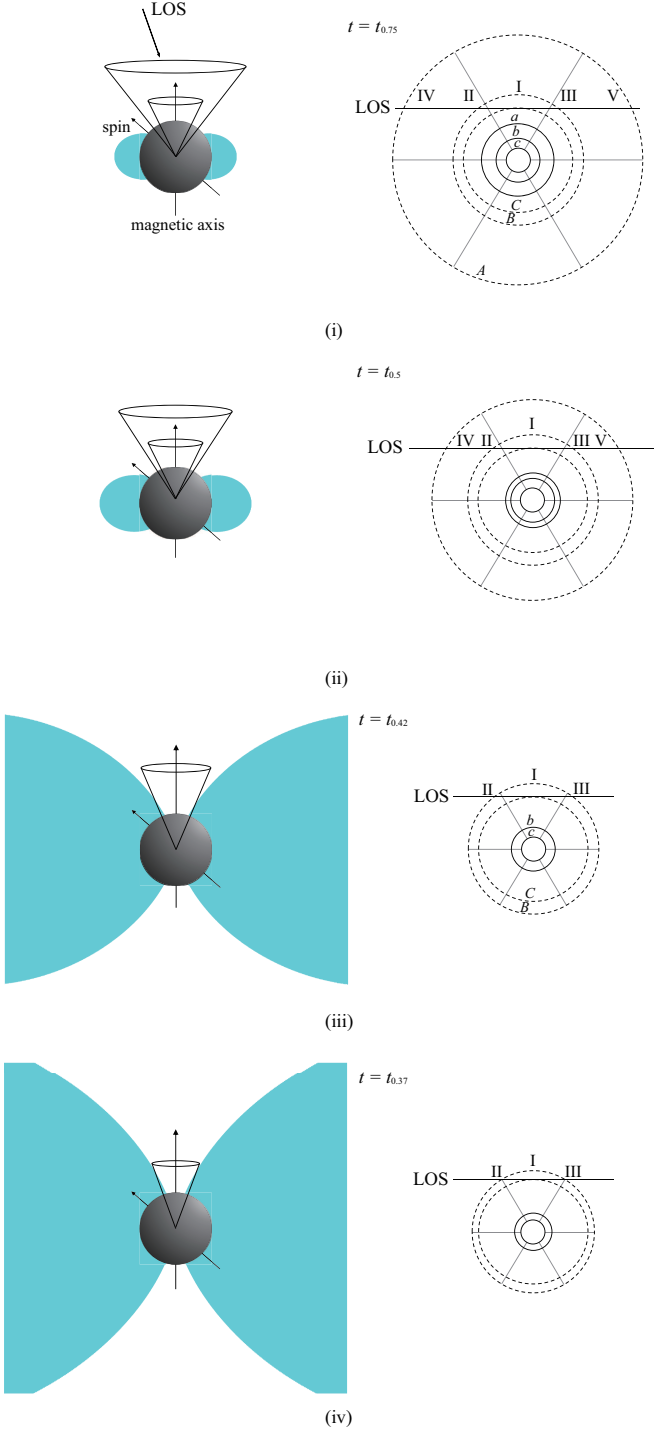}
\caption{\small{The evolution of the twisted magnetosphere (left) and the geometric configurations of the emission (right). From top to down: (i)  $t_{0.75}=t(u_b=0.75)$; (ii) $t_{0.5}=t(u_b=0.5)$; (iii) $t_{0.42}=t(u_b=0.42)$; (iv) $t_{0.37}=t(u_b=0.37)$. In the left panel, the light blue region is the u-bundle, where the field lines are untwisted. In the right panel, the phase locations of the five emission components are shown without the aberration and retardation effects, in the form of two cones (dashed rings A and B) around one central core (dashed ring C). Gray lines are the magnetic field lines. Solid rings are the charge formation annuli for each radiation cone or core. The line of sight (LOS) is marked as horizontal line in each panel.}}
\label{fig4}
\end{center}
\end{figure}

\begin{figure*}
\begin{center}
\includegraphics[width=0.75\textwidth]{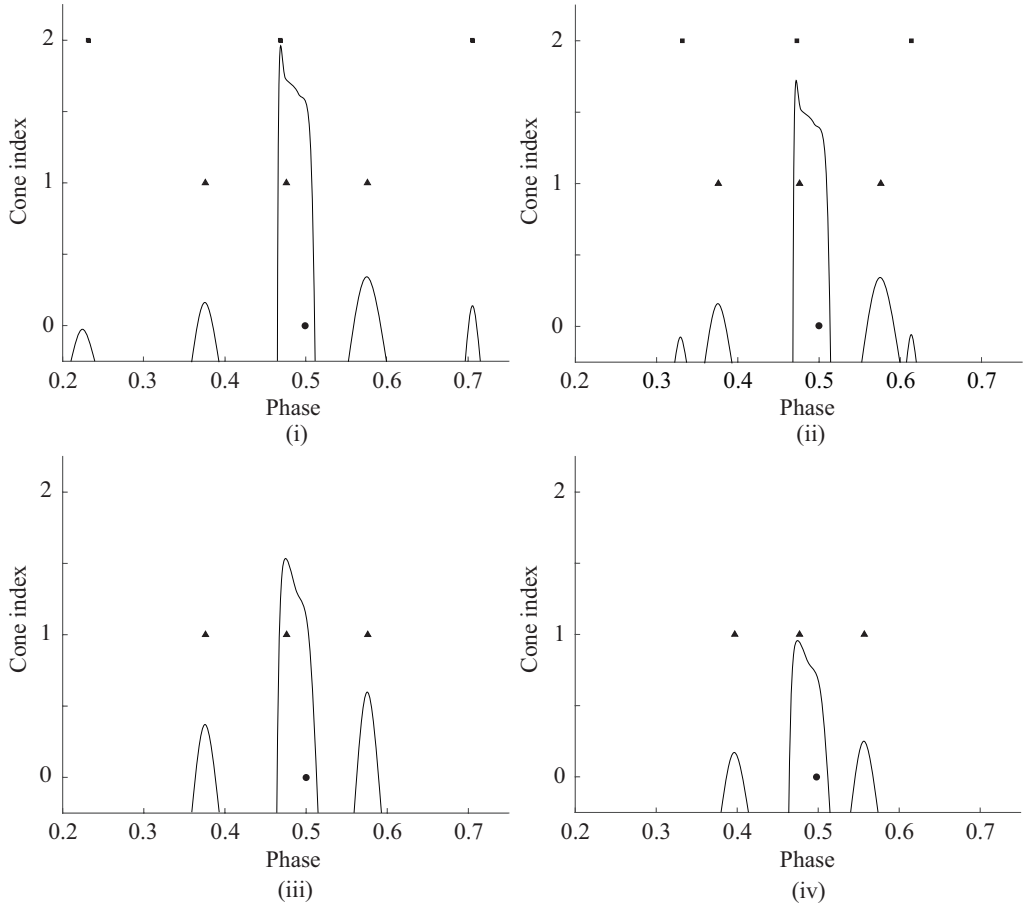}
\caption{\small{ The phases of each component during the j-bundle shrinkage and the schematics of the pulse profile evolution.} The core component is labeled as cone 0 (black round) while the inner and outer cones are labeled as cone 1 (black triangles) and cone 2 (black squares), respectively.}
\label{fig5}
\end{center}
\end{figure*}

\begin{figure}
\begin{center}
\includegraphics[width=0.48\textwidth]{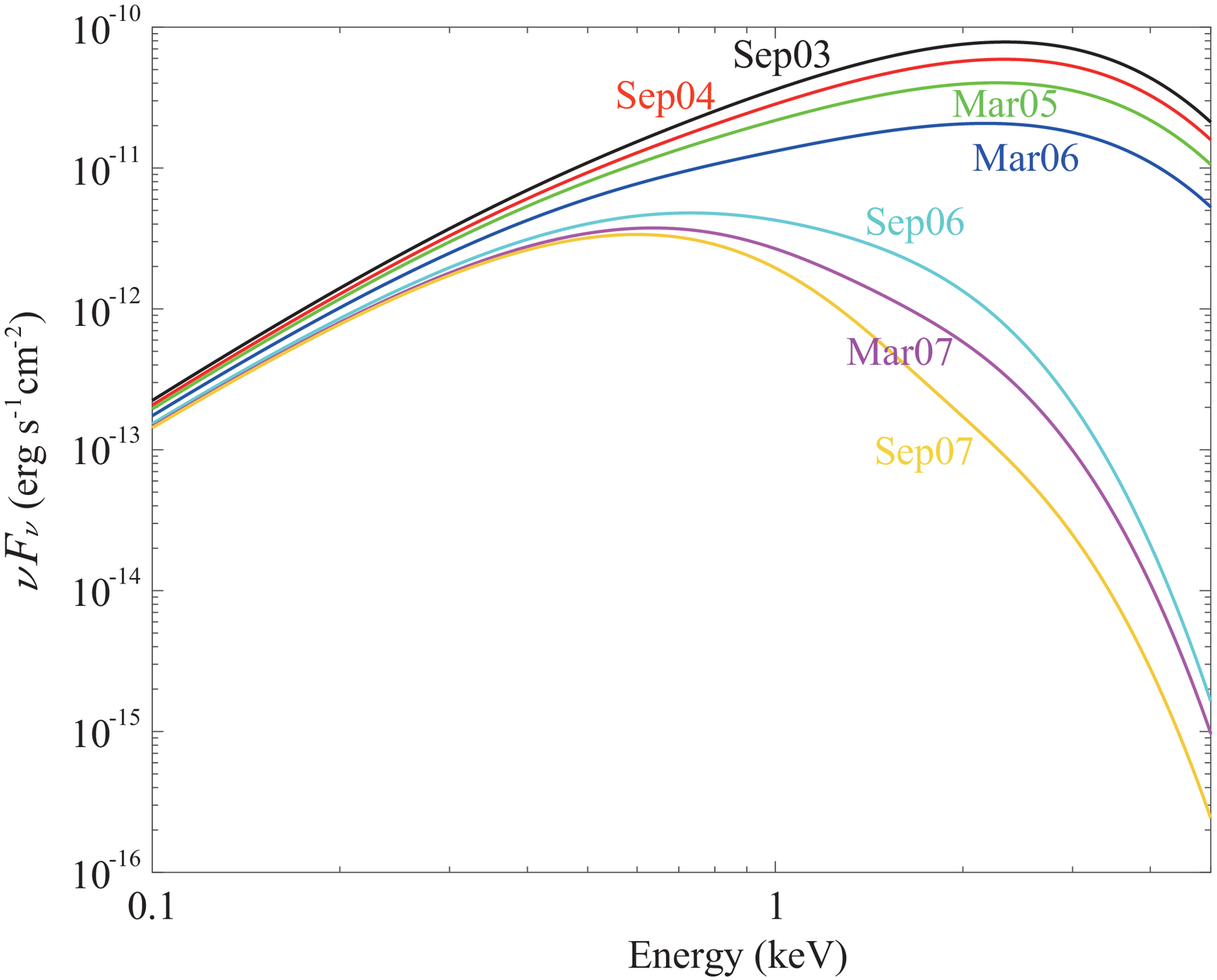}
\caption{\small{X-ray spectra of XTE 1810$-$197 after the 2003 outburst. A three-blackbody model is adopted in each spectrum. The source distance is adopted as 3.5 kpc \citep{min08}.}}
\label{fig6}
\end{center}
\end{figure}

It is widely believed that pair production is the necessary condition for pulsar radio emission \citep[e.g.][]{rud75,zhang00b}.
For an ordinary radio pulsar with $B\sim10^{12}$ G, seed electrons are accelerated to an ultra-relativistic energy with $\gamma\sim10^7$, emitting $\gamma-$rays via curvature radiation or inverse Compton scattering, which can be convert to pairs in strong magnetic fields \citep{rud75,daugherty96,zhang00}. 
The environment of a magnetar is also relevant for pair creation. 
The standard one-photon ($1\gamma$)-pair production can be suppressed due to magnetic photon splitting when $B$ becomes comparable to or exceeds $B_{\rm QED}\simeq4.4\times10^{13}$\,G \citep{bar98}. 
Photon splitting, even it is a third-order process, can compete with $1\gamma$-pair production in magnetospheres, where photons are below pair creation threshold at their emission
points \citep{har06}.
Strong vacuum dispersion may arise, so that all three CP-conservation photon-splitting modes, i.e., $\perp\to\parallel\parallel$, $\parallel\to\perp\parallel$, $\perp\to\perp\perp$, may operate together, in the special environment of a magnetar \citep{bar01}.
Consequently, photons may split before reaching the $1\gamma$-pair creation threshold.

Nonetheless, two-photon($2\gamma$)-pair production can be an important source for pair production in a magnetar environment \citep{zhang01}.
The interaction is between hard $\gamma$-rays and the copious X-ray photons from the magnetar surface. 
This process occurs mostly in the j-bundle where $E_{\parallel}$ exists.
The threshold condition for the $2\gamma$-pair production when $\epsilon_1\ll\epsilon_2$ is \citep{gou67}
\be
\epsilon_1\epsilon_2(1-\cos\beta_1)\geq2,
\label{eq19}
\ee
where $\beta_1$ is the angle between the two interacting photons, and $\epsilon_1$ and $\epsilon_2$ are the soft and hard photon energies in units of $m_{\rm e}c^2$, respectively.
For a typical blackbody radiation with temperature $kT_{B}$, the required $\gamma$-ray photon energy is 
$\epsilon_2\geq2\times10^3(kT_B/0.5\,{\rm keV})^{-1}$.

To generate such hard $\gamma$-ray photons, we consider the resonant scattering effect between an electron and an ambient X-ray photon \citep{bel07}.
In the electron rest frame, the resonant scattering happens when the blue-shifted photon frequency matches the cyclotron frequency.
The scattered photon acquires energy from Landau levels of electrons, and has a high energy $\gtrsim\gamma^2$ keV, which indicates a threshold voltage of $\gtrsim10^9$ V in the j-bundle \citep{zhang96}.
At the $\gamma$-ray energy $\epsilon_2=2\times10^3$, the attenuation length of $2\gamma$-pair production can be estimated as $\lambda_{\rm 2\gamma}=9.0\times10^3\,\rm{cm}$ \citep{gou67}.
For the same energy, the attenuation length of $1\gamma$-pair production and photon splitting can be approximated as \citep{bar01}
\be
\begin{split}
\lambda_{\rm 1\gamma}=9.2\times10^5\left(\frac{P}{1\,{\rm s}}\right)^{1/2}\left(\frac{r}{10\,\rm{km}}\right)^{1/2}\left(\frac{\theta}{0.1}\right)^{-1}\,{\rm cm},\\
\lambda_{\rm sp}=5.9\times10^4\left(\frac{P}{1\,{\rm s}}\right)^{3/7}\left(\frac{r}{10\,\rm{km}}\right)^{3/7}\left(\frac{\theta}{0.1}\right)^{-6/7}\,{\rm cm},
\label{eq20}
\end{split}
\ee
where $P$ is the spin period.
Thus, we have $\lambda_{\rm 2\gamma}<\lambda_{\rm sp}<\lambda_{\rm 1\gamma}$ \citep{zhang01}.
It means that even if $1\gamma$-pair production is suppressed by photon splitting, the $2\gamma$-pair production process can still proceed to produce copious pairs. The attenuation lengths of each process, with different $\gamma$-ray photon energies, are plotted in Figure \ref{fig3}.

Starquake-induced oscillations can drive a charge deviation from the G-J density.
This deviation triggers an electric field $E_{\parallel}$ parallel to $\boldsymbol{B}$.
The longitudinal voltage along one magnetic field line controls the ohmically released power.
To generate $2\gamma$-pair production, this voltage should not be much higher than $10^9$ V.
The pair plasma is accelerated by this voltage, maintaining the current until it is dissipated via the ohmic effect.
The $E_{\parallel}$ can be screened by pairs, but can grow again due to the continued crust oscillations, 
and the discharging repeats as the pair plasma leaves the discharging region.

\subsection{Coherent radio emission}
Theoretical models of NS coherent radio emission invoke one of three mechanisms: emission by bunches; a reactive instability, and a kinetic instability \citep{mel17}. In this paper, we consider the bunch model.
To generate coherent emission, it is required that photons are emitted in phase.
In view of the observed GHz pulse duration, the curvature radiation time scale $T_p\sim1$\,ns is much shorter than that of the observed pulse emission $T_{\rm pul}$, so that there must be more than one bunch sweeping cross the line of sight \citep{yang17}.
The half-wavelength for the 1 GHz wave is 15\,cm.
If the bunch scale is smaller than the half-wavelength, the phase of emission radiated by each particle in the bunch would be approximately the same.
The $N$ electrons in the bunch are assumed to move along nearly identical orbits, so that they act like a single macro-charge which emits a power $N^2$ times the power emitted by a single electron \citep{mel17}.
Consequently, the large fluctuating net charge of $n_{\rm GJ}$ can contribute to the coherent radiation \citep{yang17}.

We assume that the oscillation-driven charges in the j-bundle generate coherent radio emission.
It is unclear which modes would be excited and be dominant.
Here we consider a typical mode with $l = 2, m = 0$ as an illustration.
Basically, modes of $m>1$ are suppressed because $Y_{lm}$ has a term of $\sin^m\theta$ which is extremely small for $\theta\ll1$.
From equation (\ref{eqA}), the displacement amplitude is $A(R)=2.2\times10^3$ cm, if a quake energy $E_{q}=5\times10^{41}$ erg is adopted for a typical neutron star.
Note that an electric current may flow in closed field lines, which is different from the case of an ordinary radio pulsar.
However,  in the closed field lines with $r\ll R_{\rm LC}$, the unstable net charges are difficult to create and some significant absorption may exist because of the large value of the charge density there.
Hence, the coherent radio emission is suggested to be originated from the regions far from the stellar surface. 
The curvature radius $R_{c}\sim R_{\rm LC}\simeq2.4\times10^{10}(P/5\,{\rm s})\,\rm{cm}$ is adopted at these locations.
For $\theta<0.5$, from the equation (\ref{eqa1}), the radiation central position can be estimated as $r\simeq0.75R_{\rm LC}\sin\theta\simeq1.8\times10^{10}(P/5\,{\rm s})\sin\theta\,\rm{cm}$ (see Appendix).

It is also assumed that the charges obey a power-law distribution with a spectral index $p$ and energy cut-off at $\gamma_1$.
The Lorentz factor of particles accelerated by the threshold voltage is $\gamma_1=eV/(m_ec^2)\sim10^3(V/10^9\rm\,V)$.
Therefore, the electron number in the bunch volume $V_e\simeq0.1Lr^2\theta\Delta\phi\sin\theta$ is
\begin{equation}
\begin{split}
N=\frac{p-1}{\gamma_1}\frac{\delta \rho}{e}V_e\\
=4.1\times10^{13}\left(\frac{\Omega_{\rm osc}}{1\,\rm{kHz}}\right)\left(\frac{B}{10^{15}\,\rm{G}}\right)\left(\frac{L}{10\,\rm{cm}}\right)\\
\times\left(\frac{\Delta \phi}{0.1}\right)\left(\frac{\theta}{0.01}\right)\left(\frac{P}{5\,{\rm s}}\right)^{-1}\left(\frac{V}{10^9\,\rm{V}}\right)^{-1},
\label{eq14}
\end{split}
\end{equation}
where $p=2$ is adopted (see Appendix).
The observed peak flux is
\begin{equation}
\begin{split}
F_{\nu,\rm{max}}=\frac{C(p)e^2}{6\pi c}\frac{N^2\gamma_1^4}{D^2L}\left(\frac{\sin\Delta\phi}{\Delta\phi}\right)^2\left(\frac{\nu_{\rm peak}}{\nu_{c}}\right)^{2/3}\\
=21.7\left(\frac{B}{10^{15}\,\rm{G}}\right)^2\left(\frac{\Omega_{\rm osc}}{1\,\rm{kHz}}\right)^2\left(\frac{V}{10^9\,\rm{V}}\right)^{2}\left(\frac{L}{10\,\rm{cm}}\right)^2\\
\times\left(\frac{\Delta \phi}{0.1}\right)^2\left(\frac{\theta}{0.01}\right)^2\left(\frac{P}{5\,{\rm s}}\right)^{-2}\left(\frac{D}{5\,\rm{kpc}}\right)^{-2}\,\rm{mJy},
\label{eq15}
\end{split}
\end{equation}
where $D$ is the source distance, and $\nu_{\rm peak}/\nu_{c}={\rm min}[(4R_c)/(3L\gamma_1^3), 4740/(\gamma_1^3\theta^3), 1]$ (see Appendix).
This is generally consistent with the observed peak flux of a magnetar.

In this scenario, the radiation spectrum is a broken power law \citep{yang17}.
With the adopted parameters, the break frequencies in the spectrum can be calculated as $\nu_l=1.0\,{\rm GHz},\nu_{\phi}=1.4\,{\rm GHz},\nu_{c}=0.3\,{\rm GHz}$, and the broken power law spectrum is shown in Figure \ref{A2} in Appendix.
The radio spectra of the magnetars are very flat so that the higher frequency signals can be observed.
This flat spectrum is related to a power-law electron distribution with $p\approx2$. 
For instance, the galactic center magnetar PSR J1745$-$2900 shows a shallow spectral index of $-0.4\pm0.1$ at $2.54-225\,$GHz \citep{tor15}.
As shown in Figure \ref{A2}, $\nu_{\phi}>225\,$GHz is a necessary condition.
The spectrum is flat in the range of $\nu_l<\nu<\nu_c$.
From equation (\ref{afrequency}), one can calculate $\Delta\alpha<3\times10^{-4}$, if the curvature radius is roughly the light cylinder radius.
Thus, the emission region is suggested to be at near the light cylinder.
Also, in this scenario, the bunch scale is estimated to be 0.04 cm so that the millimeter waves can be coherent.
At these bands, pulse scattering and dispersion caused by interstellar medium can be neglected, which is very helpful for pulsar detection.

\subsection{Conal beam geometry}

Observationally, the radio emission of magnetars has a variety of pulse profiles and often includes multiple emission pulse components. 
In our model, we adopt a phenomenological conal beam radiative model (see \citealt{ran93}, for a review).
It has been proposed that the conal and core structures can be created via curvature radiation (e.g., \citealt{gil90,gan04}).
We assume  a circular emission beam for each conal/core component. 
Let $(r_i, \theta_i)$ be the coordinates of the emission point for the $i$th cone.
The angle $\Gamma_i$ between the pulsar magnetic axis and the magnetic field line tangential direction is calculated at the points of the $i$th cone by \citep{gil84,tho91}
\be
\sin^2\left(\frac{\Gamma_i}{2}\right)=\sin\zeta\sin\alpha\sin^2\left(\frac{\Delta\phi}{4}\right)+\sin^2\left(\frac{\zeta-\alpha}{2}\right),
\label{eq16}
\ee
where $\zeta$ is the angle between the line of sight (LOS) and the spin axis, $\alpha$ is the magnetic inclination angle, and $\Delta\phi$ is the corresponding apparent pulse width (or separation) resulting from such a geometric configuration. 
The relationship between $\Gamma_i$ and the emission point is \citep{gan01}
\be
\tan\theta_i=-\frac{3}{2\tan\Gamma_i}\pm\sqrt{2+\left(\frac{3}{2\tan\Gamma_i}\right)^2}.
\label{eq17}
\ee
If one ignores the aberration and retardation effects, 
photons are emitted tangentially along the field lines and their frequencies are determined by the curvature radius and the Lorentz factor.
The location of the emission point is determined by equation (\ref{eqa1}).
Typically, we assume that there are five emission components which consist of one core and two conal rings.

A shift of the position of conal components with respect to the core component, a.k.a. the so-called aberration and retardation effects, have been observed in many radio pulsars (see \citealt{krz09} for a review).
These effects address the bending of the radiation beam and the different paths of radiation from the conal emission regions to the observer.
The aberration and retardation always play an important roles in low frequency emission, which is believed to originate from the high-altitude regions, i.e., the regions that are far from the stellar surface.
The small net phase shift due to aberration and retardation for the $i$th cone is given by \citep{gan01}
\be
\eta_i\simeq\frac{(1+\sin\zeta)r_i}{2\pi R_{\rm LC}},
\label{eq18}
\ee
where $\zeta$ is the angle between the line of sight and the spin axis. 
Both the aberration and retardation effects make the pulse components to appear at earlier longitudinal phases.

\section{Case studies: XTE J1810$-$197 and PSR J1622$-$4950}

In the following, we apply the physical and geometric radio emission model to two magnetars and interpret their radio emission. 

\subsection{Observational properties of XTE J1810$-$197}

XTE J1810$-$197 is an AXP with spin period $P=5.54$ s and surface magnetic field $B\simeq3\times10^{14}$ G \citep{got07}.
An outburst occurred some time between 2002 November 17 and 2003 January 23 \citep{ibr04}, and the X-ray luminosity decays on a timescale of years (e.g., \citealt{hal05a}).
This object switches on as a radio pulsar during the decay of the X-rays, and shows a hard spectrum, strong linear polarization and variable pulse profiles that are very different from ordinary radio pulsars \citep{hal05b}.
From radio and X-ray observations, the geometry of XTE J1810$-$197 is inferred as $(\alpha,\,\zeta)=(52^\circ,\,29^\circ)$ \citep{ber11}, roughly consistent with the geometry we assumed in Section.3.2.

\cite{cam16} proposed that there are five peaks on the radio pulse profile of XTE J1810$-$197.
P1 appeared at the very beginning and disappeared last.
This peak was suggested to be the component IV of the outer cone in our model.
The component V is missing, perhaps because the line of sight does not sweep through the beam.
The phase of P3 does not drift, indicating that it may be the core component I.
P2 appeared at the late stage of evolution. 
Its presence in the early stage was not positively confirmed.
The interval phase of P2 and P5 get smaller because the inner cone shrinks as the u-bundle expands. 
Also, there may be some spectral evolutions which lead to their flux evolution.
These peaks may be the inner cone components II and III.
However, P4 is very close to P5 and sometimes they are mixed.
It is hard to identify P4 because there may be some multi-peak structures caused by noises.

\subsection{Radio emission modeling of XTE J1810$-$197}

We assume the initial center of the inner cone is at $r_1=0.1R_{\rm LC}$ (subscript $i=1$ for the inner cone and $i=2$ for the outer cone).
The curvature radius at 1.4 GHz is estimated as $R_{c}\simeq1.9\times10^9(\gamma/700)^3$ cm.
For simplicity, we assume that each emitting component has the same curvature radius.
Therefore, we have the angular position of the inner cone $\theta_1=0.35$.
It is worth noting that pair production sharply ends at the surface of $B\approx10^{13}$\,G \citep{bel13}.
Multipolar magnetic fields may exist near the stellar surface, leading to possible multi-hollow structures of charge formation regions.
 Emitting charges come from different formation regions which are thought to be ring-like, i.e., multiple annuli, leading to multi-cones (e.g., \citealt{gil00}), shown in Figure \ref{fig4}.
 As the u-bundle expands, the shrinkage of annulus $a$ leads to the phases of the outer cone to creep towards the core component.
The shape and the location of the annuli $b$ and $c$ are kept constant.
The twist-maintained current is intensified, for the j-bundle twists up within the frame of $dV/du>0$ (see Section \ref{sec4.3}).
From equation (\ref{eq12}), we have $dE_{\parallel}/dt>0$ and $dV/dt>0$.
Hence, the emission center for a given frequency moves toward the magnetic axis, while the emission cone shrinks as the u-bundle expands.
For simplicity, the angle $\Gamma_2$ is adopted as the central angular position of the j-bundle, i.e., $\Gamma_2\sim0.5\arcsin(u_b)^{0.5}$.
The center location of the cone (or core) is determined by equations (\ref{eq16}) and (\ref{eq18}).
As the emission center gets closer to the magnetic axis, the width of each component does not decrease, until the emission cone is eaten by the u-bundle.

Later, when the u-bundle covers the entire outer cone, the outer boundary of the emitting region reaches the tangent point to the line of sight.
Note that the area and location of each emission cone is determined by its charge formation region.
As a result, components IV and V disappear, and the inner cone starts to shrink.
Stages of the evolution of the emission components are shown in Figure \ref{fig4}.
The geometry is under the condition of $\alpha=50^\circ$ and $\zeta=30^\circ$.
Phase evolutions are also simulated and plotted in Figure \ref{fig5}.
Finally, the entire magnetosphere twists down and the radiation damps.

From the radio profile of XTE J1810$-$197 in late 2008, the position of emission cone is estimated to be $r_i\sim0.24R_{\rm LC}$ \citep{cam16} with $\theta_i\sim0.65$.
Thus, we have $u_i>u_{\rm LC}$.
The radio emission is created on the closed field lines of the j-bundle, which is different from the case of an ordinary radio pulsar.
For an ordinary radio pulsar, charged flow is difficult to form in closed field lines, whereas such global electric currents can be triggered in the closed field lines within the framework of a magnetar-quake.
The emission region is suggested to be far from the stellar surface because of its high opacity.
Such a high altitude gives rise to significant aberration and retardation.
Hence, the j-bundle plays the role of the open field lines within the framework of ordinary radio pulsars, i.e., providing electric currents.
It enlarges the beam size that increases the chance that the beam is swept by the LOS.

We have invoked core plus double conal emission components to interpret the radio pulse profiles of XTE 1810$-$197.
The outer cone is related to the boundary between the twisted (j-bundle) and untwisted (u-bundle) regions, whereas the inner cone keeps constant before the outer cone is eaten by the u-bundle.
Within the pulsar model, the core-double-cone structure was also interpreted within the framework of the inverse Compton scattering model \citep{qia98}, in which radiative particles can be generated from one annulus.
In this scenario, the shrinkage of the u-bundle leads to that the core and two cones shrink together, so that the core component may diminish because the ring C is separated from LOS. This seems not applicable to the observations of XTE 1810$-$197.
Alternatively, the patchy beam model \citep{lyn88} has been applied to interpret the pulse profiles of radio pulsars. 
This model predicts no frequency dependence for the relative pulse phase between surpluses, which is inconsistent with the systematic variation of pulse components as observed in XTE 1810$-$197. 
Finally, a fan beam ``patchy'' model was proposed by \cite{wang14} to interpret radio pulsar beams. 
It predicts that the pulse width increases with the absolute value of the impact angle, in contrast with the trend for the conal beam model. 
This can be tested by future observations with a larger sample of radio magnetars.

The decay rate of the radio flux is large from the early state of the transient because of the high ohmic dissipation rate.
It then decreases because the u-bundle expansion slows down.
The spin-down rate decreases during this period of time \citep{cam07a} because of the growth of the untwisted region and tends to be a constant later as the outflow dissipates.
This scenario broadly fits the observed spin-down behavior of XTE J1810$-$197.
In addition, some continued large-amplitude day-to-day fluctuations on the flux density of XTE J1810$-$197 are found after 2007.
These fluctuations may be caused by some aftershocks.
The quake amplitude distribution of magnetars resembles that of earthquakes, e.g. obeying the Gutenberg$-$Richter law.

\subsection{X-ray emission of XTE J1810$-$197}
\label{sec4.3}

Magnetars always show some non-thermal components in the hard X-ray band, e.g., above 20 keV, even though they have Planck-like spectra in the softer band.
The X-ray spectra of XTE J1810$-$197 after the outburst in 2003 are plotted in Figure \ref{fig6} \citep{alb10}.
The spectra can be well fitted by a three-component blackbody model.
It is found that the area of the cold component, which comes from stellar cooling, is getting larger.
In the case of twisted magnetosphere, after the starquake, the outflow particles maintain the j-bundle, where the accelerating electric field draws the positrons back from the upper pair formation front. 
These positrons fall onto the stellar surface, forming a hot spot emitting thermal X-ray photons (e.g., \citealt{har01}), which shrinks as the untwisted u-bundle expands.

Under the condition of $V(u)\approx$ const, the twist-maintained currents dissipate rapidly from the very beginning and then the decay rate decreases when $u\ll1$.
Therefore,  equation (\ref{eq6}) determines the evolution timescale of the j-bundle shrinkage.
For $V(u)$ = const, one can calculate the timescale as \citep[e.g.][]{bel09}
\be
\label{eqtime}
\begin{split}
t_{\rm ev}&=\frac{BR^2\Delta\psi_0}{cV}\\
&\simeq 15\left(\frac{\Delta\psi_0}{0.1}\right)\left(\frac{B}{10^{15}\,\rm{G}}\right)\left(\frac{R}{10\,\rm{km}}\right)^2\left(\frac{V}{10^{9}\,\rm{V}}\right)^{-1}\,\rm{yr}.
\end{split}
\ee
This is generally consistent with the observed timescale of radio luminosity decay (e.g., \citealt{cam16}).
The twist angle is also a constant until it is eaten by the expanding u-bundle front.
Additionally, the free energy stored in the twisted magnetosphere is $E_{\rm tw}\simeq B^2R^3(\Delta\psi)^2/24\sim4\times10^{44} (B/10^{15}\,{\rm G})^2(R/10\,{\rm km})^3(\Delta\psi_0/0.1)^2\,{\rm erg}$ \citep{bel09}.
One can calculate an average luminosity
\be
\begin{split}
L_{X}\sim \frac{E_{\rm tw}}{t_{\rm ev}}=2.6\times10^{36}\left(\frac{B}{10^{15}\,{\rm G}}\right)\\
\left(\frac{R}{10\,{\rm km}}\right)\left(\frac{\Delta\psi_0}{0.1}\right)\left(\frac{V}{10^9\,{\rm V}}\right)\,{\rm erg\,s^{-1}}.
\label{eq21}
\end{split}
\ee

Consider the condition of $dV/du>0$ in the j-bundle.
As discussed in Section.3.1, in principle, a decreased energy of hard $\gamma$-ray photons near the magnetic axis would imply a negative pair number gradient $dn_{\rm pair}/du$.
Therefore, the hot spot heating rate decreases as the colatitude increases.
One can also obtain $d\Delta\psi/dt>0$ and $dj/dt>0$ at the j-bundle \citep{bel09}.
This does not result in an extra net charge fluctuation because the evolution timescale is much longer than the duration of a pulse.
A fraction of the released energy during the u-bundle twisting down propagates into the j-bundle, so that the j-bundle magnetic fields twist up.
This may be the reason why the observed luminosity one year after the 2003 outburst is below the estimated luminosity $L_{X}$.

\subsection{The case of PSR J1622$-$4950}
PSR J1622$-$4950 is another radio emitting magnetar with a period of $P=4.3$ s and nearly $100\%$ linear polarization \citep{lev10}.
It has a flat spectrum, highly variable flux density and pulse profiles during 2009 to 2011, with the X-ray flux decreasing by an order of magnitude \citep{and12}.
Detectable radio emission was observed from 1999 to 2003, and from 2011 November to 2014 March \citep{sch17}.
The radio flux decreases (ranging over $\sim3-80$ mJy) and finally disappeared as the entire magnetosphere is untwisted.
The peak flux during the evolution is a few to several tens mJy, similar to that observed in XTE 1810$-$197, matching the calculation of equation (\ref{eq15}).

However, the pulse profile of this object consists of two main bright peaks, so only one emission cone is needed without the need of introducing a core component.
From equation (\ref{eq16}) and (\ref{eq17}), if one assumes $\alpha=20^\circ$ and $\zeta=10^\circ$ (e.g., \citealt{lev12}), the position of emission cone is estimated as $r_i\sim0.1R_{\rm LC}$ with $\theta_i=0.71$.
Therefore, one has $u_i\sim 5u_{\rm LC}$.
The two components are emitted from the closed field lines of the j-bundle.
These peaks tend to get closer \citep{sch17} since the emission cone shrinks while the u-bundle expands.
The timescale of the pulse profile variation is similar to that of the flux density and the torque,  which meets the estimation from equation (\ref{eqtime}).
The variable torque of the magnetar, as a common property for magnetars following outbursts, was also observed in this object \citep{cam18}.

The X-ray evolution also shares similar properties with the case of XTE J1810$-$197.
However, the X-ray observations are not sufficient to indicate that the area of the hot spot is getting smaller.
Also, no X-ray outburst was detected before its radio decay.
From the exponential decay of the X-ray flux during 2007-2011, \cite{and12} argued that an undetected X-ray outburst occurred not long before mid-2007.
Possibly, another X-ray outburst would have also occurred not long before 1999.

\section{Summary and Discussion}

A model of coherent curvature radiation for magnetars is proposed within the scenario of a magnetar-quake-induced, twisted, and oscillating magnetosphere.
We show that radio emission originates from the twisted current-carrying bundle (the j-bundle), which is similar to the open field line region of normal pulsars (cf. \citealt{bel09}, who suggested that the radio emission is from the closed field line region).
Continued oscillations are excited in this region due to the aftershocks and untwisting of the magnetosphere, with pairs generated via two-photon processes. 
The j-bundle shrinks with time as the untwisted region (the u-bundle) increases, so that the radio emission beam progressively becomes narrower. 
The radio emission profile evolves with time as a consequence of the shrinkage of the j-bundle, and disappears when the radio emission beam is small enough to escape the LOS. 
This model predicts a peak flux of radio emission and a flat spectrum that are generally consistent with the observations of magnetar radio emission. 
The shrinkage of the j-bundle is also consistent with the evolution of the X-ray hot spot of the magnetars during the radio active phase.
We apply a time-dependent conal-beam radiative model to successfully explain the variable radio pulsation behaviors of XTE J1810$-$197 and PSR J1622$-$4950.

Our study shows that magnetars most likely have a different radio emission mechanism from ordinary radio pulsars. 
Even though the coherent mechanism is similar (bunched curvature radiation), the mechanisms to excite bunches are different. 
Whereas radio pulsars likely trigger bunches through unsteady pair production from the polar cap region defined by the open field lines, magnetars trigger bunches through quake-driven oscillations and continued untwisting of the magnetosphere. 
The twisted magnetosphere serves as an effective open field line region, which shrinks as a function of time. 
The shrinkage of this effective open field line region is the ultimate reason for the transient nature of magnetar radio emission. 

The difference of the emission mechanism between magnetars and normal pulsars is also reflected on their pair production mechanisms.
PSR J1119$-$6127, for instance, is a highly magnetized radio pulsar.
\cite{arc16} found that the persistent X-ray flux increased by a factor of 160 with a large glitch following the X-ray bursts.
Unlike radio transient magnetars, the radio emission quenches following an X-ray burst, and reappears roughly two weeks later \citep{bur16a,bur16b}.
Even the magnetic field is very high for this object, $1\gamma$-pair production may not be suppressed.
The X-ray burst may have formed a fireball, making pair plasma density exceeding the G-J density by orders of magnitude.
The leakage of these pairs to the polar cap region may have screened the parallel electric field and quenched the radio emission \citep{arc17}. 
This is different from the magnetar case, because the triggering mechanism for pair production and radio emission is very different for magnetars.

\acknowledgments
{{We are grateful to Jiguang Lu and Yuan Shi at National Astronomical Observatories, Chinese Academy of Sciences, Hao Tong and Hongguang Wang at Guangzhou University, all of the members in the pulsar group at Peking University, and an anonymous referee for helpful comments and discussions about the model. W.Y.W. and X.L.C. acknowledge the support of MoST 2016YFE0100300, Natural Science Foundation of China (NSFC) 11473044, 11633004, 11653003, CAS QYZDJ-SSW-SLH017. R.X.X. acknowledges the support of National Key R\&D Program of China (No. 2017YFA0402602), NSFC 11673002 and U1531243, and the Strategic Priority Research Program of CAS (No. XDB23010200).}}

\appendix
\section{Dipolar geometry}
We make some brief calculations of the pulsar geometry. For a dipole magnetic field, the curvature radius is 
\be
R_c=\frac{[r^2+(dr/d\theta)]^{3/2}}{|r^2+2(dr/d\theta)^2- rd^2r/d\theta^2|}=\frac{r(1+3\cos^2\theta)^{3/2}}{3\sin\theta(1+\cos^2\theta)}.
\label{eqa1}
\ee
It can be reduced to that of a simple relationship when $\theta\ll1$, i.e.,
\be
R_c=\frac{4r}{3\sin\theta}.
\ee
The angle between the magnetic axis and the magnetic field is 
\be
\alpha=\theta+\arccos\left(\frac{2\cos\theta}{\sqrt{1+3\cos^2\theta}}\right).
\ee
For $\theta\ll1$, its derivation is
\be
\frac{d\alpha}{d\theta}=\frac{3(1+\cos^2\theta)}{1+3\cos^2\theta}\approx\frac{3}{2}.
\ee
The bunch volume can be estimated as \citep{yang17}
\be
V\simeq\frac{2}{3}Lr^2\sin\theta\Delta\alpha\Delta\phi,
\ee
where $\Delta\alpha$ and $\Delta\phi$ are the bunch opening angles.

\section{Coherent radio emisssion}
\begin{figure}
\begin{center}
\includegraphics[width=0.48\textwidth]{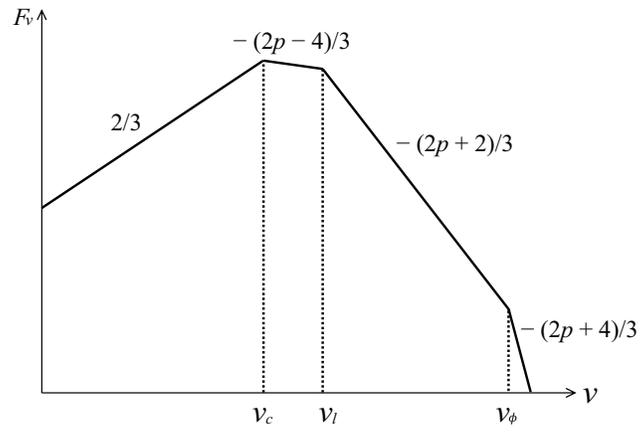}
\caption{\small{A multi-segment broken power law emission spectrum for coherent curvature radiation by bunches. }}
\label{A2}
\end{center}
\end{figure}

For a single electron, the energy radiated per unit frequency interval per unit solid angle is given be \citep{jack98}
\be
\frac{dI}{d\omega d\Omega}=\frac{e^2\omega^2}{4\pi^2c}(|A_{\parallel}|^2+|A_{\perp}|^2),
\ee
where two polarized components of amplitude are
\be
\begin{split}
A_{\parallel}=\frac{2iR_c^{1/3}3^{1/6}}{c^{1/3}}\left(\frac{\xi}{\omega}\right)^{2/3}K_{2/3}(\xi),\\
A_{\perp}=\frac{2R_c^{2/3}\theta}{3^{1/6}c^{2/3}}\left(\frac{\xi}{\omega}\right)^{1/3}K_{1/3}(\xi),
\end{split}
\ee
where $\xi=\omega R_c(1/\gamma^2+\theta^2)^{3/2}/(3c)$ and $K(\xi)$ is the modified Bessel function.
For electrons with a power-law of $N_e=N(\gamma/\gamma_1)^{-p}$ from $\gamma_1<\gamma<\gamma_2$, a coherent sum of amplitudes is
\be
\begin{split}
\left(\frac{dI}{d\omega d\Omega}\right)_{\rm coherent}=\frac{e^2\omega^2}{4\pi^2c}\times\\
\left(\left|\int^{\gamma_2}_{\gamma_1}N_e(\gamma)A_{\parallel}(\omega,\gamma)d\gamma\right|^2+\left|\int^{\gamma_2}_{\gamma_1}N_e(\gamma)A_{\perp}(\omega,\gamma)d\gamma\right|^2\right).
\end{split}
\ee
The normalization of electron distribution is $N=(p-1)\gamma_1^{-1}\delta nV$.
In this picture, the spectrum is a broken power-law \citep{yang17}. The break frequencies are
\be
\nu_{c}=\frac{3c\gamma_1^3}{4\pi R_c}, ~~~ \nu_l=\frac{c}{\pi L}, ~~~ \nu_{\phi}=\frac{12c}{\pi R_c(\Delta \alpha)^3}.
\label{afrequency}
\ee
For the given parameters in Section.3.1, the spectrum are plotted in Figure \ref{A2}.
The peak flux is
\be
\begin{split}
F_{\nu, {\rm max}}=\frac{2\pi}{TD^2}\left(\frac{dI}{d\omega d\Omega}\right)_{\rm coherent}\\
\simeq\frac{C(p)e^2}{6\pi c^2}\frac{N^2\gamma_1^4}{D^2T}\left(\frac{\sin\Delta\phi}{\Delta\phi}\right)^2\left(\frac{\nu_{\rm peak}}{\nu_{c}}\right)^{2/3},
\end{split}
\ee
where $C(p)=4^{p/3}[\Gamma(2/3)\Gamma((p-1)/3)]^2$ and $T$ is the observation time.

\end{document}